\documentclass[runningheads]{llncs}

\usepackage[T1]{fontenc}
\usepackage{amsmath,amssymb}
\usepackage{array}
\usepackage{booktabs}
\usepackage{float}
\usepackage{hyperref}
\usepackage{jlcode}
\usepackage{makecell}
\usepackage[capitalise]{cleveref}
\usepackage{graphicx}

\usepackage{tikz}
\usetikzlibrary{arrows.meta,decorations.pathreplacing,calc}

\colorlet{bestcolor}{green!15}

\newcommand{\QQ}{\mathbb{Q}}
\newcommand{\ZZ}{\mathbb{Z}}
\newcommand{\grobner}{Gr\"obner}
\newcommand{\code}[1]{\texttt{#1}}

\begin{document}

\title{Groebner.jl: Fast Gr\"obner Tracing in Julia}
\titlerunning{Groebner.jl}

\author{Alexander Demin\thanks{
This work has been supported by an ERC-2023-ADG grant for the ODELIX project (number 101142171) and partially supported by the NSF grant CCF-2212460.
\\[0.6em]
\noindent\begin{minipage}{0.72\linewidth}
{\it Funded by the European Union. Views and opinions expressed are however those of the author(s) only and do not necessarily reflect those of the European Union or the European Research Council Executive Agency. Neither the European Union nor the granting authority can be held responsible for them.}
\end{minipage}
\hfill
\begin{minipage}{0.24\linewidth}
\centering
\protect\includegraphics[height=1.1cm]{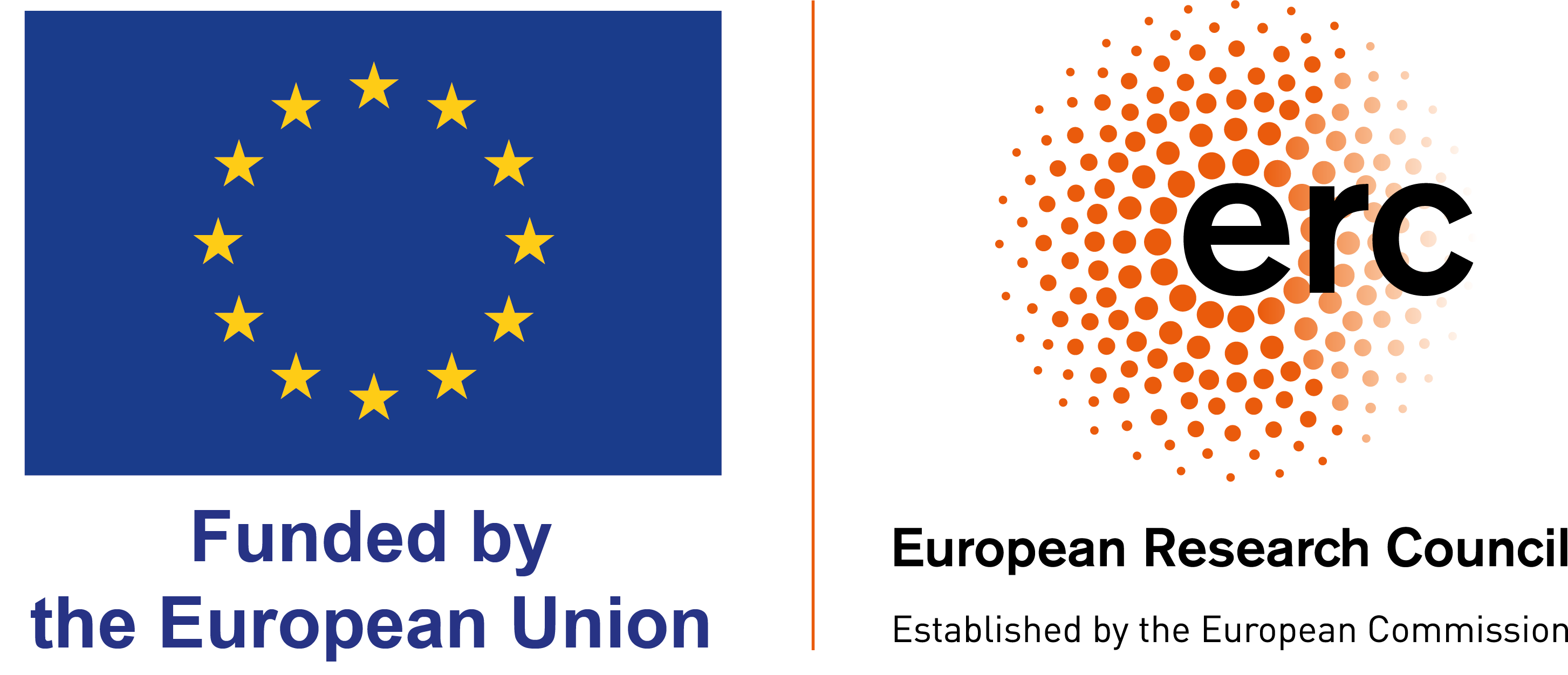}
\end{minipage}
}
}
\authorrunning{Demin}

\institute{
Laboratoire d'informatique de l'{\'E}cole polytechnique,\\LIX, UMR 7161, CNRS,\\1 rue Honor{\'e} d'Estienne d'Orves,\\91120 Palaiseau, France\\
\email{demin@lix.polytechnique.fr}}

\maketitle

\begin{abstract}
A standard way to control expression swell in computer algebra is to use multi-modular or evaluation-interpolation methods. In computations involving \grobner{} bases, these techniques typically require repeatedly computing \grobner{} bases of specializations of the same ideal. These repeated computations can be accelerated through precomputation, notably using Traverso’s tracing. 

We present \href{https://github.com/sumiya11/Groebner.jl}{\code{Groebner.jl}}, a Julia implementation of the F4 algorithm that exposes Traverso's tracing through a reusable public interface. 
The implementation supports SIMD-friendly coefficient types, such as tuples of machine integers, which Julia compiles to efficient code with little manual intervention.
This lets other Julia software leverage tracing to obtain speedups in applications such as structural identifiability of ordinary differential equation models and polynomial system solving.

\end{abstract}

\section{Introduction}

When using multi-modular and evaluation-interpolation methods, it is typical to repeatedly compute \grobner{} bases of specializations of the same ideal. Traverso's tracing~\cite{tracingtrav} is a method of precomputation that accelerates such repeated calculations.
One first performs a \emph{learn} computation, a modified \grobner{} basis computation that additionally records which S-polynomials reduce to zero.
One may then perform several \emph{apply} computations on specializations of the same ideal, omitting reductions of S-polynomials that are expected to reduce to zero.

High-performance implementations of the F4 algorithm~\cite{F4} use tracing internally for computing \grobner{} bases over the rationals, including \code{msolve}~\cite{msolve,kouba2026complexityanalysisf4grobner} and Giac/Xcas~\cite{giac}; related ideas appear in~\cite{joux-vitse}. Typically, these systems do not expose tracing through their user interfaces. Many applications, however, use \grobner{} basis computations as intermediate subroutines rather than final outputs. Therefore, providing a reusable learn \& apply interface would make tracing available to a broader class of symbolic computation workflows.

We discuss \code{Groebner.jl}, a Julia package that implements the F4 algorithm and exposes Traverso's tracing through public learn and apply functions~\cite{gowda-demin}. To further amortize computational costs, the apply routine can compute \grobner{} bases in batches. The implementation uses SIMD-friendly coefficient types and relies on Julia's LLVM-based compiler to produce efficient code across architectures with little additional implementation effort. We evaluate these techniques on applications in structural identifiability~\cite{simple_generators,SI} and polynomial system solving via rational univariate representations~\cite{demin-rouillier-ruiz}.

\paragraph{Acknowledgements.}
We would like to thank Roman Pearce, Gleb Pogudin, Fabrice Rouillier, and Chee Yap for useful discussions. We are also grateful to the anonymous referees for helpful comments and suggestions.
The project is open-source, and we are grateful to all contributors; in particular, to Matthias Franz.

\section{Gr\"obner tracing}
\label{sec:tracing}

This section briefly describes the ideas behind \grobner{} tracing using the F4 algorithm as an example. We refer to~\cite{tracingtrav} for details. Tracing consists of two stages, both of which may be viewed as modifications of the classic F4 algorithm, as illustrated in \Cref{fig:trace-iteration}:

\vspace{0.0em}
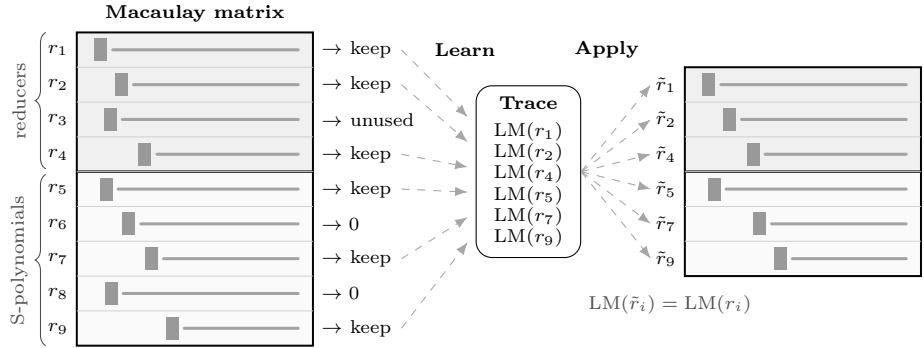
\begin{figure}[H]
\centering
\begin{tikzpicture}[x=0.39cm,y=0.46cm,>=Latex,font=\scriptsize]
    \def\LearnX{0}
    \def\MatrixW{8.0}
    
    \def\GapLearnTrace{7.3}
    \def\GapTraceApply{4.0}
    
    \pgfmathsetmacro{\TraceX}{\LearnX+\MatrixW+\GapLearnTrace}
    \pgfmathsetmacro{\ApplyX}{\TraceX+\GapTraceApply+1.3}

    \newcommand{\DrawRowAt}[5]{%
    \fill[LeadFill]
        ($(#1)+(#3,#2-0.32)$)
        rectangle ++(0.42,0.64);

    \draw[TailStroke,line width=1.15pt,line cap=round]
        ($(#1)+(#4,#2)$)
        --
        ($(#1)+(#5,#2)$);
    }

    \colorlet{LeadFill}{black!40}
    \colorlet{TailStroke}{black!36}
    \colorlet{ReducerTone}{black!6}
    \colorlet{SPolyTone}{black!2}

    \colorlet{AnnotationTone}{black!70}
    \colorlet{ArrowTone}{black!35}


    \node at (4.0,9.55) {{\bf Macaulay matrix}};
    \fill[ReducerTone] (0,5) rectangle (8.0,9);
    \fill[SPolyTone] (0,0) rectangle (8.0,5);
    \draw[line width=0.75pt] (0,0) rectangle (8.0,9);
    \draw[line width=0.9pt] (0,5) -- (8.0,5);
    \foreach \y in {1,...,8}
        \draw[black!18,line width=0.25pt] (0,\y) -- (8.0,\y);

    \foreach \lab/\y in {
        $r_1$/8.5,$r_2$/7.5,$r_3$/6.5,$r_4$/5.5,
        $r_5$/4.5,$r_6$/3.5,$r_7$/2.5,$r_8$/1.5,$r_9$/0.5}
        \node[left] at (0,\y) {\lab};

    \draw[decorate,decoration={brace,mirror,amplitude=4pt},AnnotationTone]
    (-1.05,9)--(-1.05,5.08)
    node[midway,left=10pt,rotate=90,anchor=center,AnnotationTone]{reducers};

\draw[decorate,decoration={brace,mirror,amplitude=4pt},AnnotationTone]
    (-1.05,4.92)--(-1.05,0)
    node[midway,left=10pt,rotate=90,anchor=center,AnnotationTone]{S-polynomials};

    \fill[LeadFill] (0.60,8.18) rectangle ++(0.42,0.64);
    \draw[TailStroke,line width=1.15pt,line cap=round] (1.22,8.5)--(7.50,8.5);

    \fill[LeadFill] (1.30,7.18) rectangle ++(0.42,0.64);
    \draw[TailStroke,line width=1.15pt,line cap=round] (1.92,7.5)--(7.50,7.5);

    \fill[LeadFill] (0.92,6.18) rectangle ++(0.42,0.64);
    \draw[TailStroke,line width=1.15pt,line cap=round] (1.54,6.5)--(7.50,6.5);

    \fill[LeadFill] (2.10,5.18) rectangle ++(0.42,0.64);
    \draw[TailStroke,line width=1.15pt,line cap=round] (2.72,5.5)--(7.50,5.5);

    \fill[LeadFill] (0.78,4.18) rectangle ++(0.42,0.64);
    \draw[TailStroke,line width=1.15pt,line cap=round] (1.40,4.5)--(7.50,4.5);

    \fill[LeadFill] (1.55,3.18) rectangle ++(0.42,0.64);
    \draw[TailStroke,line width=1.15pt,line cap=round] (2.17,3.5)--(7.50,3.5);

    \fill[LeadFill] (2.32,2.18) rectangle ++(0.42,0.64);
    \draw[TailStroke,line width=1.15pt,line cap=round] (2.94,2.5)--(7.50,2.5);

    \fill[LeadFill] (0.95,1.18) rectangle ++(0.42,0.64);
    \draw[TailStroke,line width=1.15pt,line cap=round] (1.57,1.5)--(7.50,1.5);

    \fill[LeadFill] (3.02,0.18) rectangle ++(0.42,0.64);
    \draw[TailStroke,line width=1.15pt,line cap=round] (3.64,0.5)--(7.50,0.5);

    \foreach \txt/\y in {
        {$\to$ keep}/8.5,
        {$\to$ keep}/7.5,
        {$\to$ unused}/6.5,
        {$\to$ keep}/5.5,
        {$\to$ keep}/4.5,
        {$\to 0$}/3.5,
        {$\to$ keep}/2.5,
        {$\to 0$}/1.5,
        {$\to$ keep}/0.5}
        \node[anchor=west] at (8.0,\y) {\txt};


    \node[
        draw,
        rounded corners=6pt,
        fill=white,
        align=center,
        inner sep=4pt,
        text width=1.1cm
    ] (tracebox) at (\TraceX,5.0)
    {
        \textbf{Trace}\\[2pt]
        $\mathrm{LM}(r_1)$\\
        $\mathrm{LM}(r_2)$\\
        $\mathrm{LM}(r_4)$\\
        $\mathrm{LM}(r_5)$\\
        $\mathrm{LM}(r_7)$\\
        $\mathrm{LM}(r_9)$
    };

    \coordinate (lmrone)   at ([xshift=-0.8cm,yshift= 0.72cm]tracebox.center);
    \coordinate (lmrtwo)   at ([xshift=-0.8cm,yshift= 0.39cm]tracebox.center);
    \coordinate (lmrfour)  at ([xshift=-0.8cm,yshift= 0.06cm]tracebox.center);
    \coordinate (lmsone)   at ([xshift=-0.8cm,yshift=-0.27cm]tracebox.center);
    \coordinate (lmsthree) at ([xshift=-0.8cm,yshift=-0.60cm]tracebox.center);
    \coordinate (lmsfive)  at ([xshift=-0.8cm,yshift=-0.93cm]tracebox.center);


    \draw[dashed,->,ArrowTone,line width=0.4pt]
    (11.0,8.5) -- (lmrone);

    \draw[dashed,->,ArrowTone,line width=0.4pt]
        (11.0,7.5) -- (lmrtwo);
    
    \draw[dashed,->,ArrowTone,line width=0.4pt]
        (11.0,5.5) -- (lmrfour);
    
    \draw[dashed,->,ArrowTone,line width=0.4pt]
        (11.0,4.5) -- (lmsone);
    
    \draw[dashed,->,ArrowTone,line width=0.4pt]
        (11.0,2.5) -- (lmsthree);
    
    \draw[dashed,->,ArrowTone,line width=0.4pt]
        (11.0,0.5) -- (lmsfive);

    \node[align=center]
    at ({(11.0+\TraceX)/2},8.5)
    {{\bf Learn}};


    \fill[ReducerTone] (\ApplyX,5)
    rectangle ({\ApplyX+\MatrixW},8);

    \fill[SPolyTone] (\ApplyX,2)
    rectangle ({\ApplyX+\MatrixW},5);

    \draw[line width=0.75pt]
    (\ApplyX,2)
    rectangle
    ({\ApplyX+\MatrixW},8);
    
    \draw[line width=0.9pt] (\ApplyX,5) -- ({\ApplyX+\MatrixW},5);

    \foreach \y in {3,...,7}
        \draw[black!18,line width=0.25pt]
            (\ApplyX,\y)--({\ApplyX+\MatrixW},\y);

    \foreach \lab/\y in {
        $\tilde{r}_1$/7.5,
        $\tilde{r}_2$/6.5,
        $\tilde{r}_4$/5.5,
        $\tilde{r}_5$/4.5,
        $\tilde{r}_7$/3.5,
        $\tilde{r}_9$/2.5}
        \node[left] at (\ApplyX,\y) {\lab};

    \fill[LeadFill] ({\ApplyX+0.60},7.18) rectangle ++(0.42,0.64);
\draw[TailStroke,line width=1.15pt,line cap=round]
    ({\ApplyX+1.22},7.5)--({\ApplyX+7.50},7.5);

    \fill[LeadFill] ({\ApplyX+1.30},6.18) rectangle ++(0.42,0.64);
\draw[TailStroke,line width=1.15pt,line cap=round]
    ({\ApplyX+1.92},6.5)--({\ApplyX+7.50},6.5);

\fill[LeadFill] ({\ApplyX+2.10},5.18) rectangle ++(0.42,0.64);
\draw[TailStroke,line width=1.15pt,line cap=round]
    ({\ApplyX+2.72},5.5)--({\ApplyX+7.50},5.5);

\fill[LeadFill] ({\ApplyX+0.78},4.18) rectangle ++(0.42,0.64);
\draw[TailStroke,line width=1.15pt,line cap=round]
    ({\ApplyX+1.40},4.5)--({\ApplyX+7.50},4.5);

\fill[LeadFill] ({\ApplyX+2.32},3.18) rectangle ++(0.42,0.64);
\draw[TailStroke,line width=1.15pt,line cap=round]
    ({\ApplyX+2.94},3.5)--({\ApplyX+7.50},3.5);

\fill[LeadFill] ({\ApplyX+3.02},2.18) rectangle ++(0.42,0.64);
\draw[TailStroke,line width=1.15pt,line cap=round]
    ({\ApplyX+3.64},2.5)--({\ApplyX+7.50},2.5);


    \draw[dashed,->,ArrowTone,line width=0.4pt]
        (tracebox.east) -- (\ApplyX-1.15,7.5);

    \draw[dashed,->,ArrowTone,line width=0.4pt]
        (tracebox.east) -- (\ApplyX-1.15,6.5);

    \draw[dashed,->,ArrowTone,line width=0.4pt]
        (tracebox.east) -- (\ApplyX-1.15,5.5);

    \draw[dashed,->,ArrowTone,line width=0.4pt]
        (tracebox.east) -- (\ApplyX-1.15,4.5);

    \draw[dashed,->,ArrowTone,line width=0.4pt]
        (tracebox.east) -- (\ApplyX-1.15,3.5);

    \draw[dashed,->,ArrowTone,line width=0.4pt]
        (tracebox.east) -- (\ApplyX-1.15,2.5);

    \node[align=center]
    at ({(\TraceX+\ApplyX)/2},8.5)
    {{\bf Apply}};

\node[
    text=AnnotationTone,
    align=left
]
at ({\ApplyX-0.5},1.2)
{$\mathrm{LM}(\tilde r_i)=\mathrm{LM}(r_i)$};

\end{tikzpicture}%
\caption{A single iteration of the F4 algorithm as a row-reduction of a Macaulay matrix. Learn stage (on the left) and apply stage (on the right). $\mathrm{LM}$ denotes the leading monomial without coefficient.}
\label{fig:trace-iteration}
\end{figure}

\begin{itemize}
    \item[Learn stage.] {\bf Input:} a system of polynomials $F_1$. {\bf Output:} a trace. The algorithm performs a modified F4 computation on $F_1$. At each iteration, it constructs and row-reduces a Macaulay matrix. In addition to the operations of the standard F4 algorithm, it records which rows of the matrix reduce to zero, and stores this information in a coefficient-independent manner in a trace.

    \item[Apply stage.] {\bf Input:} a system of polynomials $F_2$ and a trace. {\bf Output:} a \grobner{} basis of $F_2$. The algorithm uses the trace to construct only the rows of the Macaulay matrix that were identified as useful during the learn stage. As a result, it avoids many of the monomial operations performed during learning and may produce a smaller matrix. It then row-reduces the matrix.
\end{itemize}

Tracing is applicable whenever several \grobner{} basis computations share some monomial structure but differ in their coefficients. In this setting, $F_1$ and $F_2$ are typically obtained as specializations of a common polynomial system $F$. 

\begin{example}
Let $F \subseteq \mathbb{F}_p(a)[x_1,\ldots,x_n]$. To compute the reduced \grobner{} basis of $F$, one may first specialize $a$ to obtain a system $F_1$ and perform the learn stage. The resulting trace can then be reused for many further specializations $F_2,F_3,\ldots$, each requiring only the apply stage. The \grobner{} bases of these specializations can subsequently be passed to rational function interpolation to recover the basis over $\mathbb{F}_p(a)$.
\end{example}

\section{Interface design}
\label{sec:interface}

\subsection{Generic coefficient arithmetic}

\code{Groebner.jl} computes Gr\"obner bases of polynomial ideals over fields.
It can be used in combination with existing symbolic manipulation packages, such as \code{Nemo.jl}, or by itself through a low-level interface.
A standard workflow is:

\begin{lstlisting}[columns=fullflexible]
using Groebner, Nemo                    # load packages

K, t = rational_function_field(QQ, "t") # define Q(t)[x,y,z]
R, (x, y, z) = K["x","y","z"]

F = [t*x^2*y + y^3, x*y^2 + 3//t*z]     # define polynomials

G = groebner(F)                         # compute basis
\end{lstlisting}
This produces output of the following form:
\begin{lstlisting}
4-element Vector{AbstractAlgebra.Generic.MPoly{...}}:
 y^6 + 9//t*z^2
 x*z - 1//3*y^4
 x*y^2 + 3//t*z
 x^2*y + 1//t*y^3
\end{lstlisting}

The implementation in \code{Groebner.jl} is generic with respect to coefficient types. 
Thus, the package can operate over fields for which the basic arithmetic operations are implemented, whether in \code{Groebner.jl} itself or in external packages. In the example above, arithmetic in $\QQ(t)$ is supplied by \code{Nemo.jl}. For certain coefficient types, \code{Groebner.jl} provides more efficient specialized implementations. Such specializations only need to implement coefficient arithmetic and, optionally, a number of performance-critical primitives, while the rest of the implementation is reused unchanged. For example, the implementation over integers modulo a prime $p$ with $p < 2^{64}$ uses machine integers together with specialized routines for linear algebra and vector operations modulo $p$. 

\subsection{Learn and Apply}

\code{Groebner.jl} exports tracing via the learn and apply interface.
We illustrate it by computing \grobner{} bases for two specializations of the Katsura-10 system:

\begin{lstlisting}[columns=fullflexible]
kat = Groebner.Examples.katsuran(10)

# use Nemo to produce two copies of the system modulo different primes
kat_zp1 = change_base_ring.(Ref(GF(2^30+3)), kat)
kat_zp2 = change_base_ring.(Ref(GF(2^30+7)), kat)

trace, _ = groebner_learn(kat_zp1);   # learn a trace for the first prime

success, gb = groebner_apply!(trace, kat_zp2);  # apply for another prime
\end{lstlisting}
The boolean \code{success} reports whether any inconsistencies were detected by comparing the supports of the polynomials against the trace.
This test may produce false positives: no inconsistency may be detected even when the result is not a \grobner{} basis.
In practice, however, it can still serve as a useful proxy.

The learn and apply pattern is particularly appealing when many applications are required, since each application is faster than an independent \grobner{} basis computation. 
To illustrate this, we compare the running time of the default implementation in \code{Groebner.jl} against that of an application:

\begin{lstlisting}
using BenchmarkTools

@btime groebner(kat_zp2);
#  731.000 ms (84551 allocations: 139.38 MiB)

@btime groebner_apply!(trace, kat_zp2);
#  263.576 ms (44002 allocations: 87.02 MiB)
\end{lstlisting}

\begin{remark}[Choice of baseline]\label{remark:baseline}
The default implementation in \code{groebner} uses the randomized linear algebra techniques following~\cite{magma,compactF4}, which is state of the art. Consequently, it serves as the baseline throughout this paper. For comparison, the classical deterministic implementation takes $3.5$ seconds on this example.
\end{remark}

\subsection{Application in batches}\label{sec:batches1}

The application interface supports computing several \grobner{} bases simultaneously. For example, one may perform an application with a batch of four systems, all obtained as specializations of the same system:
\begin{lstlisting}
# ... define sys1,...,sys5 appropriately ...
trace, _ = groebner_learn(sys1)
success, (gb2,gb3,gb4,gb5) = groebner_apply!(trace,(sys2,sys3,sys4,sys5))
\end{lstlisting}
The output corresponds to the \grobner{} bases of the systems \code{sys2}, \code{sys3}, \code{sys4}, and \code{sys5}, respectively. Internally, this allows certain operations to be shared across the four computations (see~\Cref{sec:batches2}). For the Katsura-10 example, the application in batches of four takes approximately 680 ms; this corresponds to 170 ms per \grobner{} basis, compared to 264 ms for a single application.

\section{Efficient \grobner{} tracing}
\label{sec:design}

\subsection{When is tracing worthwhile?}
\label{sec:worthwhile}

The ``learning'' of a trace is typically more expensive than a single classic \grobner{} basis computation. A natural question is therefore: how many \grobner{} bases do we need to compute before this cost is amortized?

\begin{table}[H]
\centering
\setlength{\tabcolsep}{4pt}
\caption{Running times, and the number of computations required to amortize the cost of learning a trace compared against classic independent \grobner{} basis computations.}
\begin{tabular}{lrrrr}
System & $\mathrm{Classic}$ & $\mathrm{Learn}$ & $\mathrm{Apply}$ & Computations to break even \\
\midrule
Chandra-11 & 1.0 s & 3.3 s & 0.1 s & 4 \\
Chandra-12 & 5.8 s & 19.6 s & 0.3 s & 4 \\
Katsura-12 & 2.2 s & 17.2 s & 0.8 s & 13 \\
Katsura-13 & 14.2 s & 153.2 s & 6.1 s & 19 \\
Cyclic-8 & 0.4 s & 1.5 s & 0.2 s & 8 \\
Cyclic-9 & 38.9 s & 219.4 s & 26.7 s & 19 \\
Cholera & 13.7 s & 40.9 s & 7.7 s & 7 \\
Yang1 & 5.8 s & 39.2 s & 0.4 s & 8 \\
\end{tabular}
\label{tab:breakeven}
\end{table}

In~\Cref{tab:breakeven}, for each benchmark system, we present the running time of a standard independent \grobner{} basis computation (per~\Cref{remark:baseline}), together with the running times of the learn and apply stages. Using these timings, we compute the smallest number of \grobner{} basis computations for which learning a trace once and subsequently applying it repeatedly is faster than performing all computations independently. The resulting number is shown in the last column.

The first thing we observe is that the learn stage is always slower than the classic computation, and the apply stage is always faster.
We also note that the break-even point typically is small. Consequently, when hundreds or more \grobner{} bases of specializations are required (which is typical for multi-modular and evaluation-interpolation methods), the learn \& apply strategy may be beneficial.

The table also shows that the benefit of tracing depends on the problem. For example, the ratio of running times $\frac{\mathrm{Classic}}{\mathrm{Apply}}$ is only $1.46$ for Cyclic-9 but reaches $14.5$ for Yang1. In Yang1, a substantial fraction of the running time in the classic algorithm is spent on monomial operations, many of which are eliminated during the apply stage. In contrast, Cyclic-9 spends a larger fraction of its running time in linear algebra, which must still be performed during application, albeit on smaller matrices.

\subsection{Application in batches}
\label{sec:batches2}

Instead of computing Gr\"obner bases over integers modulo primes $p_1, \dots, p_N$ independently, one may compute over the product ring
\begin{equation}\label{eq:product-ring}
\ZZ/p_1\ZZ \times \cdots \times \ZZ/p_N\ZZ,
\end{equation}
whose elements are represented as tuples of residues with arithmetic performed component-wise. As a result, operations in the F4 algorithm such as critical-pair processing and monomial arithmetic are performed only once, shared across the primes. Thus, increasing $N$ amortizes the cost of this part of the computation.
This approach has been employed by Gr\"abe~\cite{Graebe1993LuckyPrimes} and de Kleine and Monagan~\cite{dekleine-monagan-batch}.

\code{Groebner.jl} can compute over the product rings~\eqref{eq:product-ring} for primes $p_i < 2^{64}$ with $i=1,\ldots,N$. The application in batches from~\Cref{sec:batches1} is implemented in this way. Since the F4 implementation is generic with respect to coefficient arithmetic, only a small number of linear algebra kernels need to be specialized for the arithmetic in~\eqref{eq:product-ring}, while the remainder of the learn \& apply infrastructure is reused unchanged.

As an illustration of one such kernel, consider addition of a dense vector and a multiple of a sparse vector modulo a prime, one of the bottlenecks of the classic F4 algorithm. 
Consider implementing this operation over~\eqref{eq:product-ring} for $N=4$ using $31$-bit primes each represented in \code{Int64} and using the modular arithmetic from~\cite{compactF4}. With the broadcast syntax in Julia, such as \code{.+}, \code{.*}, \code{.-}, etc.\footnote{For example, in Julia, \code{(1,2,3,4) .+ (5,6,7,8) == (6,8,10,12)}}, this can be implemented as simply as (cf.~\cite[Section 2.2]{compactF4}):

\begin{lstlisting}[columns=fullflexible]
const N = 4
const p = (2^30+3, 2^30+7, 2^30+9, 2^30+15) # p_1, p_2, p_3, p_4
const ProductRingElem = NTuple{N,Int64}

function reduce(
        buffer::Vector{ProductRingElem},    # dense vector
        coeffs::Vector{ProductRingElem},    # sparse vector coeffs.
        idxs::Vector{Int64})                # sparse vector support
    p2 = p .^ 2
    c = buffer[idxs[1]]
    for k in 1:length(idxs)
        j = idxs[k]
        a = buffer[j] .- c .* coeffs[k]
        buffer[j] = a .+ ((a .>> 63) .& p2) # (a < 0) ? a + p2 : a
    end
end
\end{lstlisting}
This is essentially the implementation used in \code{Groebner.jl}. Apart from the broadcast operations, it is identical to the scalar implementation for $N=1$.

In addition to the classical benefits of product-ring arithmetic, the resulting operations are particularly amenable to SIMD vectorization~\cite{joris-modular}; this contrasts with the scalar case, where vectorization is difficult because of irregular memory access patterns and the low arithmetic intensity in sparse linear algebra. For component-wise tuple operations in product rings we found that Julia, via LLVM, often automatically generates SIMD-optimized assembly. On Intel processors, this leads to substantial amortized speedups; on ARM architectures such as Apple M2 Max, the generated code was less effective in our experiments\footnote{It is possible to inspect the assembly of the \code{reduce} function using the command:\\ {\scriptsize \code{code\_native(reduce, Tuple\{Vector\{ProductRingElem\},Vector\{ProductRingElem\},Vector\{Int64\}\})}}}.

\begin{remark}[The case $p_1=\cdots=p_N$]
When $p_1=\cdots=p_N$, the computation may be viewed as performing $N$ independent computations modulo the same prime simultaneously. This arises in evaluation-interpolation schemes, where one computes \grobner{} bases for many specializations of a parametric polynomial system over a fixed finite field.
\end{remark}

\subsection{The choice of batch size}

Computing over the product ring~\eqref{eq:product-ring} introduces a parameter $N$, the number of primes processed simultaneously. As discussed in~\Cref{sec:batches2}, operations such as critical-pair processing and monomial arithmetic are shared across all components of the product ring. Increasing $N$ therefore amortizes the cost of these operations. On the other hand, larger values of $N$ increase the amount of coefficient data and may increase memory traffic and reduce cache efficiency.

To study this trade-off, we measure the throughput of the apply stage for different values of $N$. For each benchmark system and each value of $N$, \Cref{table:pick-N} reports the amortized speedup relative to the scalar implementation with $N=1$.

\begin{table}[H]
\centering
\setlength{\tabcolsep}{4pt}
\caption{The amortized speed-up of using the apply stage in batches over the scalar case $N=1$. {\bf Boldface} marks the best speed-up in a row.}
\begin{tabular}{lrrrrrr}
System & $N=1$ & $N=2$ & $N=4$ & $N=8$ & $N=16$ & $N=32$\\
\midrule
Chandra-12 & 1.00 & 1.30 & 1.58 & 1.87 & {\bf 2.51} & 2.05 \\
Chandra-13 & 1.00 & 1.39 & 1.93 & 2.05 & {\bf 2.27} & 1.81\\
Katsura-12 & 1.00 & 1.50 & 2.23 & {\bf 2.33} & 2.19 & 2.12\\
Katsura-13 & 1.00 & 1.45 & 2.04 & 2.11 & 2.24 & {\bf 2.26}\\
Cyclic-8 & 1.00 & 1.69 & 2.79 & 3.14 & {\bf 3.28} & 2.68\\
Cyclic-9 & 1.00 & 1.50 & 2.09 & 2.19 & {\bf 2.20} & 2.06 \\
Cholera & 1.00 & 1.46 & 1.94 & 1.96 & 1.98 & {\bf 2.00}\\
Yang1 & 1.00 & 1.04 & 1.19 & {\bf 1.24} & 1.14 & 0.95\\
\end{tabular}
\label{table:pick-N}
\end{table}

The results show that batching consistently improves throughput. The best speedups range from $1.24$ to $3.28$. We also observe that the gains from increasing $N$ saturate quickly: values between $8$ and $16$ often provide the best performance, while larger values rarely yield much additional improvement.

The Yang1 system again provides an instructive example. As discussed in~\Cref{sec:worthwhile}, tracing eliminates a substantial amount of monomial operations in this benchmark, which have been the bottleneck. Since batching amortizes the same type of work, its additional benefit is comparatively modest.

\begin{remark}[Memory consumption]
Increasing $N$ also increases memory consumption. For Katsura-13, the maximum RAM usage grows from $2.7$ GB at $N=1$ to $29.3$ GB at $N=32$. As the throughput appears to saturate well before this point, large values of $N$ are unattractive. In particular, \code{Groebner.jl} internally uses $N=4$ for multi-modular computation over the rationals.
\end{remark}

\section{Applications}

In this paper, we are using {\tt Groebner.jl} version 0.10.3. All experiments we carried out on an Intel Core i9-13900 running Linux with 128 GB of RAM; the CPU does not support AVX512.

\subsection{Polynomial system solving}

One of the standard approaches to solving zero-dimensional polynomial systems over the rationals is via rational univariate representations~\cite{rouillier99}.
In modern implementations, the computation is typically performed using multi-modular techniques, requiring many \grobner{} basis computations modulo different primes.
Our experiments use \code{Rational}\allowbreak\code{Univariate}\allowbreak\code{Representation}\allowbreak\code{.jl}, which implements the algorithms from~\cite{demin-rouillier-ruiz}.

\begin{table}[H]
\centering
\small
\setlength{\tabcolsep}{5pt}
\caption{Total running times of rational univariate representation computation.}
\begin{tabular}{lrrrr}
System & \thead{Backend:\\\code{groebner}} & \thead{Backend:\\\code{groebner\_apply}} & \thead{Backend:\\\code{groebner\_apply} ($N=4$)} & Speedup \\
\midrule
Chandra-10 & 82.4 s & 32.5 s & 29.3 s & 2.8\\
Chandra-11 & 851.4 s & 395.5 s & 373.5 s & 2.3\\
Eco-10 & 1.0 s & 0.6 s & 0.4 s & 2.5\\
Eco-11 & 11.2 s & 6.5 s & 4.2 s & 2.7\\
Katsura-11 & 71.0 s & 39.8 s & 28.2 s & 2.5\\
SEIR36 & 63.7 s & 11.2 s & 9.6 s & 6.6\\
cp\_d3\_n6\_p6 & 66.4 s & 49.0 s & 35.9 s & 1.8\\
\end{tabular}
\label{table:rur}
\end{table}

We compare three methods on several benchmark systems.
The first uses \code{groebner} for all \grobner{} basis computations.
The second learns a trace from the first modular computation and then reuses it for the remaining primes.
The third, shown in the \code{groebner\_apply} ($N=4$) column, uses the application in batches from~\Cref{sec:batches1} to process four primes at once during the apply stage.

The results show that both tracing and batching can substantially accelerate solving, with the overall speedups range from $1.8$ to $6.6$.

\subsection{Identifiable functions of dynamical models}

Structural identifiability is the problem of determining whether the parameters ${\boldsymbol \mu} = (\mu_1,\ldots,\mu_m)$ of a dynamical model can be recovered from the observed data.
For models defined by ordinary differential equations, an important task is to compute generators of the field of identifiable functions, a subfield of $\QQ(\boldsymbol \mu)$.

The algorithm of~\cite{simple_generators}, implemented in \code{StructuralIdentifiability.jl}~\cite{SI}, computes such generators by interpolating selected coefficients of the reduced \grobner{} basis of an ideal in $\QQ(\boldsymbol \mu)[\mathbf{x}]$.
The workflow repeatedly specializes the parameters ${\boldsymbol \mu}$ and reduces modulo primes, producing a sequence of \grobner{} basis computations in $\ZZ/p\ZZ[x]$ with fixed monomial structure but varying coefficients. 
The learn-and-apply strategy is therefore a natural fit.

\begin{table}[H]
\centering
\small
\setlength{\tabcolsep}{5pt}
\caption{The number of evaluation points and the total running times for computing generators of the field of identifiable functions via rational function interpolation using \code{groebner} and \code{groebner\_apply}.}
\begin{tabular}{lrrrr}
Model & \thead{Evaluation points} & \thead{Backend:\\\code{groebner}} & \thead{Backend:\\\code{groebner\_apply}} & Speedup \\
\midrule
EAIHRD~\cite{Fokas2020} & 420 & 300 s & 49 s & 6.1\\
LinComp2~\cite{ahmed2025identifiabilitydirectedcyclecatenarylinear} & 4388 & 275 s & 164 s & 1.7\\
Pharm~\cite{Demignot1987Pharm} & 34 & 54 s & 36 s & 1.5\\
\end{tabular}
\label{table:ident}
\end{table}

We consider the models EAIHRD~\cite{Fokas2020}, LinComp2~\cite{ahmed2025identifiabilitydirectedcyclecatenarylinear}, and Pharm~\cite{Demignot1987Pharm}.
For each model, we run \code{StructuralIdentifiability.jl} and compare a workflow using only \code{groebner} with the one combining \code{groebner\_learn} and \code{groebner\_apply}; see~\Cref{table:ident}.
The results confirm that tracing can accelerate this interpolation-based workflow.

\bibliographystyle{splncs04}
\bibliography{refs}

\end{document}